**OPEN FORUM**

# Security practices in AI development

Petr Spelda[1] · Vit Stritecky[1]

## Abstract

What makes safety claims about general purpose AI systems such as large language models trustworthy? We show that rather than the capabilities of security tools such as alignment and red teaming procedures, it is security practices based on these tools that contributed to reconfiguring the image of AI safety and made the claims acceptable. After showing what causes the gap between the capabilities of security tools and the desired safety guarantees, we critically investigate how AI security practices attempt to fill the gap and identify several shortcomings in diversity and participation. We found that these security practices are part of securitization processes aiming to support (commercial) development of general purpose AI systems whose trustworthiness can only be imperfectly tested instead of guaranteed. We conclude by offering several improvements to the current AI security practices.

## 1 Introduction

Safety of foundation AI models (Bommasani 2021), such as large language models capable of in-context learning (Brown et al. 2020), is based on aligning a pretrained model with human preferences on the model's responses to user inputs and then testing whether alignment with the preferences holds in as wide a range of situations as possible. The pair of an alignment method, e.g., reinforcement learning from human feedback generalizing preferences through reward modeling (Ouyang et al. 2022, Bai et al. 2022a), and a testing method, e.g., systematic adversarial testing called red teaming attempting to find gaps left behind by the alignment method (Ganguli et al. 2022; Casper et al. 2023a), represents a tool allowing the owner of the model to enact a security practice.

With every new release of commercial chatbots based on LLMs (large language model)[1], we are witnessing yet another round of enacting some security practice that aims to manage the risk in the eyes of the public and regulators alike. The practice itself enables and is accompanied by narratives of safe product development that protects consumers, society and democracy from harmful effects of AI. There is, however, a gap between the capabilities of the tools (the pairs of alignment and testing methods) and safety guarantees which are promised by enacting security practices with the tools.

We will show that since it is hard to maintain the promised guarantees with available tools, security practices that aim to fill the gap follow the logic of Didier Bigo's (in) securitization processes (Bigo and Tsoukala 2008). The (in) securitization process captures routines of risk management enacted by actors aiming to create a space in which they can make claims about (in)security (ibid.). In the sense of (in) securitization processes, AI security practices enacted by alignment and testing methods are no different than the practices designed to sustain (in)security claims about migration, health or terrorism which originally motivated the critical approaches to the study of security (e.g., C.A.S.E. Collective 2006). We argue that safety claims about LLMs and other types of foundation models are impossible to evaluate without understanding the gap between capabilities of security tools and the desired security guarantees. The gap is to be filled with enacting security practices of risk management consisting of routines designed to manage unease about or even fear of for-profit, privately-owned and closed development of capable AI models.

✉ Petr Spelda
  petr.spelda@fsv.cuni.cz

1  Department of Security Studies, Institute of Political Studies, Faculty of Social Sciences, Charles University, U Kříže 8 Praha 5, 158 00 Prague, Czech Republic

---

1  This applies to a degree also to publicly available LLMs.

In order to explain the motivation and critically assess the risks associated with security practices in LLMs, or in general foundation models, development, we proceed as follows. In Sect. 2, it will be explained what causes the gap between security tools and the desired guarantees and, thus, motivates the security practices. In Sect. 3, we will show that the security practices supporting closed, for-profit development of LLMs can cause a lack of participation, accountability, transparency and democratic legitimacy and above all cannot guarantee that the gap between capabilities of the tools and the desired guarantees will be closed. We will conclude with Sect. 4 on more open, participatory, and sustainable practices in LLM development.

## 2 AI alignment cannot guarantee AI safety

The main method for aligning capabilities of LLMs with human preferences used today is reinforcement learning from human feedback (RLHF; Ouyang et al. 2022; Bai et al. 2022a). The core of RLHF is reward modeling, a method that learns from human preferences over alternative outputs of the model that is being aligned. The aim of learning a reward model from human preferences is to generalize from them and be able to provide correct rewards for outputs of the model that were not included in the preference dataset.

Reward modeling was developed to make training of deep reinforcement learning agents in simulated environments (e.g., games, physics simulators) more efficient with respect to human feedback (Christiano et al. 2017). The reward model captured human preferences on pairs of trajectories of the agent in the environment and learned to generalize from them so that for that particular environment the job of human judges could be automated (ibid.). The same principle has been applied for aligning pretrained LLMs with human preferences on LLM outputs to sensitive inputs. Sensitive inputs include prompts for which pretrained, unaligned models are known to generate biased and inequitable responses (Tamkin et al. 2023) or a wide range of unsafe responses (Mazeika et al. 2024).

Experiments such as OpenAI's InstructGPT (Ouyang et al. 2022) utilized reward modeling based on human preferences for producing models capable of following natural language instructions in a better way compared to, for instance, the GPT-3 model that was trained using the next token prediction objective and was not explicitly aligned for instruction following. Schematically, the procedure can be described in four steps (ibid.):

1. For a set of prompts, people write responses that reflect their preferences (alternatively, suitable existing responses are collected).

2. The set of prompt-response pairs is used for supervised fine-tuning of some base model pre-trained using the next token prediction objective, providing a foundation for an instruction following model IFM.

3. People provide preferences for training the reward model. For each prompt, a ranking of alternative responses to the prompt, $A \succ C \succ B \sim D$ (last two are ranked as equal), is produced by human rankers, and the resulting set of binary preferences over alternatives (e.g., $A \succ D$ pairwise ranking) is used to train a reward model that generalizes from the preferences.

4. For a new prompt, the IFM generates a response evaluated by the reward model which produces a reward for it and the reward is used to update the IFM using a reinforcement learning method.

Often, for cost-effectiveness reasons, human rankers are replaced with existing reward models or generative models prompted to select a response from two options, which can replace reward models (see Lambert et al. 2024a, p. 17). While far from optimal considering issues such as participation or inclusion that we discuss in Sect. 4, automated preference generation allowed scaling-up LLM development (relatedly, see also Bai et al. 2022b). The inherent tradeoff should, however, not be overlooked. There is also a class direct alignment algorithms that do not rely on explicit reward models, most notably Direct Preference Optimization (DPO, Rafailov et al. 2023). DPO does not involve training a reward model from human preferences or reinforcement learning. Rather, DPO calculates the probability of preferred and dispreferred response to a prompt under the model and optimizes its parameters to increase the probability of preferred over dispreferred responses (ibid.). There is evidence (e.g., Lambert et al. 2024b) that DPO and its variants can offer strong performance at the post-training alignment stage. Post-training alignment methods can be, of course, combined. See, for instance, how Lambert et al.'s (2024b) recipe uses DPO for preference tuning and reinforcement learning with verifiable rewards, a variant of RLHF replacing the reward model with verifiers, to strengthen the model's mathematical capabilities. For the class of generative models that produce chain-of-thought traces before outputting a response, there are methods such as Deliberative Alignment (Guan et al. 2024), trying to ensure that the model's chain-of-thought trace is informed by a safety policy relevant for the user's prompt. The method uses a reward model with access to safety policies to filter high-quality completions for supervised fine-tuning and for the final, reinforcement learning stage (ibid.).

It is important to note that the RLHF-based alignment procedure described in steps 1–4 is a domain adaptation rather than safety-guaranteeing method. The method can be used to adapt a base model for instruction following.

Interpreting this as a safety procedure requires that the human preferences expressed in steps 1 and 3 allow the reinforcement learning stage to remove all unsafe capabilities of the instruction following model inherited from the base model.

This is clearly impossible to guarantee for several reasons. First, even if guidelines for providing feedback are based on an equitable and inclusive conception of safety, people are bound to provide incomplete feedback and biased, inconsistent or confused preferences accidentally, out of malice or due to a perceived difficulty of the comparison task (cf. Casper et al. 2023b). Second, it is difficult, if not impossible, to ensure that the reward model will not misgeneralize the preferences during training and will be able to serve as a good guide for updating the IFM during its alignment (ibid.). Finally, given the considerable generality of LLMs trained on the biggest piles of data ever assembled, it is impossible to cover every harmful capability.

Testing the model to uncover its unsafe capabilities, which need to be removed, cannot secure full coverage either. The impossibility can be explained by building an analogy with software testing. The main obstacle to verifying by testing that a software system possesses some properties according to a specification is what Symons and Horner (2019) call the 'path-complexity catastrophe'. Their argument relies on an observation that even a simple program logic induces exponential scaling of execution paths, which cannot be all tested for adherence with the specification due to computational intractability of such an effort (Symons and Horner 2019, Section 2.2; tests parallelization cannot solve the path-complexity catastrophe either, ibid.). If then, the software's overall reliability is inferred from a statistic calculated from the executed tests, the inference becomes an inductive prediction about the system's reliability. And because we need to make assumptions about the system's overall error distribution (which will remain unknown due to the path-complexity catastrophe, ibid.) to perform that prediction, it cannot provide unproblematically admissible evidence about the system's overall reliability.

By the same token, can we predict that an LLM will provide aligned completions to user tasks on all future occasions from a limited sample of its responses obtained from performed tests? It is possible to try, but the prediction will not lead to verifiable guarantees with respect to some alignment specification captured as human preferences or other constraints on correct answers. The problem is worse for poisoned (Rando and Tramèr 2024; Hubinger et al. 2024) or deceptively aligned models (see early toy experiments, e.g., Meinke et al. 2024) that will breach their alignment regardless of pre-deployment testing known as red teaming.

Harmful capabilities are found by red teaming (Feffer et al. 2024; Inie et al. 2023; Casper et al. 2023a). Red teaming consists of prompting techniques that aim to elicit harmful capabilities of the model and help to find the unavoidable 'gaps' left behind by the alignment procedure. In its core, red teaming is just an empirical testing practice with adversarial goals that can be performed by people as well as by other LLMs tasked with generating adversarial prompts (Perez et al. 2022; Samvelyan et al. 2024)[2]. Identification of harmful capabilities is followed by repeating the alignment procedure on prompts identified by red teaming. It is crucial to note that red teaming does not refer to a well-defined practice (Feffer et al. 2024) but rather to any testing procedure that can lead to discovering harmful capabilities of supposedly safe models. Such practices can be highly domain specific and context-dependent, which makes them open to issues of participation and fairness (Fazelpour et al. 2024) like any other social practice with safety goals. The alignment-red teaming cycle can be repeated several times before and after the model is deployed, e.g., turned into a product accessible via an API or released as a publicly available model (even if here the usual practice is pre-deployment alignment-red teaming, e.g., Touvron 2023).

Unfortunately, even several repetitions of the alignment-red teaming cycle cannot provide dependable safety guarantees. First, aligned and tested LLMs are known to be vulnerable to jailbreak attacks which force the models to respond in ways that violate their alignment (Wei et al. 2023). These attacks use specialized prompts and are hypothesized to exploit the failures of alignment generalization to unsafe capabilities that were not discovered during red teaming or exploit conflicts that can arise between general instruction following and responding safely to restricted requests (ibid.). Second, Zou et al. (2023) developed an optimization-based method for finding 'adversarial suffixes' which, when appended to a wide range of restricted requests, break the LLM's alignment and force it to suppress the refusal response and return unsafe answers. It was observed that the attack suffixes are effective against a range of different LLMs and considering the fact that the search for them is automated, they substantially reduce dependability of safety guarantees that can be derived from alignment. The safety guarantees cannot be fundamentally improved by combining alignment with filtering LLM outputs according to some safety rules. Glukhov et al. (2023) showed that the filtering problem is formally undecidable and cannot be used to fix the non-robustness of LLM alignment. Despite this ultimate formal obstacle, there is some evidence that input and output classifiers trained on synthetic data generated by LLMs according to a safety specification can make jailbreaking of LLMs harder (Sharma et al. 2025).

---

[2] Classifiers can be used to determine whether the output is harmful to allow scaling up the process.

The problem is worse for models that can be further fine-tuned after alignment-red teaming was applied, e.g., for publicly available or open-source base models or commercial models offering fine-tuning APIs. Qi et al. (2024a) showed that less than 100 samples is enough for breaking alignment of a model and recovering its harmful capabilities. The robustness of alignment can be also decreased by fine-tuning an aligned and red-teamed model on benign prompt–response pairs (ibid.). Since there is currently a rich offering of publicly available or open-source high-quality, instruction-following models whose fine-tuning is cheap and can be done on high-performance consumer hardware, alignment and red teaming provides even less safety assurances than for commercial chatbots based on LLMs. In this situation, unlearning harmful knowledge from the model combined with increasing the hardness of re-learning it is considered one of possible solutions. However, it is challenging to perform unlearning in a robust way, preventing re-learning via fine-tuning, while not degrading safe capabilities of the model (cf., Barez et al. 2025). This becomes even more problematic if dual-use knowledge is involved, which has safe, legitimate uses but can also lead to model completions that pose risks in cybersecurity or CBRN (chemical, biological, radiological and nuclear) domains (ibid.).

In sum, alignment methods and red teaming practices used today cannot deliver dependable safety and fairness guarantees because gaps in LLMs' alignment are unavoidable and their alignment can be broken or compromised in multiple ways depending on access to the model. This shows that the current alignment methods based on preference modeling are good for product development, e.g., chatbots based on instruction following LLMs, and bad for developing AI products with safety guarantees. Since policy-makers demand safety guarantees (European Union AI Act[3], Biden's U.S. Presidential Executive Order on the Safe, Secure, and Trustworthy Development and Use of Artificial Intelligence[4], rescinded on January 20, 2025) to protect the public from harmful effects of LLMs whose alignment is incomplete and non-robust, companies developing these models attempt to manage the (in)security and unease over AI development with security practices.

## 3 AI alignment and red teaming as tools of security practices

We established that the gap between the tools and the desired safety guarantees cannot be currently closed, and this means that claims about safety of LLMs by the companies that develop them need to be understood as part of a securitization process. What is most visible of this process in the public space are speech acts that aim to communicate preparedness for responsible development of more capable AI systems (Anthropic 2023a; 2023b; OpenAI 2023a; OpenAI 2023b; Leike and Sutskever 2023). The securitization process follows the logic of tacitly acknowledging the shortcomings of available alignment and safety testing methods and claiming that the issues will be addressed by broad research programs whose announcements constitute the core of the speech acts. The level of detail ranges from vague and general (Leike and Sutskever 2023) to partially concrete by pointing toward existing AI safety research (Anthropic 2023a). It is important to understand that these high-level speech acts do not exist in a vacuum. They are a product of security practices which are performed using AI alignment and red teaming methods. Since these AI safety tools are limited, it is necessary to critically assess the practices that they help to enact. This critical assessment is the first step in understanding high-level speech acts regarding AI safety that are hard to parse if kept separated from the underlying security practices.

Bigo and Tsoukala (2008) initiated the study of (in)securitization processes through the lens of security practices and Balzacq et al. (2010) provided a useful framework for understanding the tools of security practices. Despite the fact that Balzacq et al.'s (2010) tools' characteristics were derived from counterterrorism and other areas, the characteristics are fitting AI safety tools well because the enactment of practices using imperfect tools always aims to fill the gap between capabilities and the desired safety guarantees.

First, security tools have design traits and defining features that make them unique yet relatable to other tools (ibid.). As explained in Sect. 2, the core of RLHF, a widely used AI alignment method today, is a reward model based on human preferences. The reward model is the design trait of RLHF considered as a security tool and human preferences are its defining features. The first critical point regarding security practices enacted with RLHF is about defining features of reward models.

There is evidence showing that human preferences used to train reward models represent only some social groups (e.g., Santurkar et al. 2023), which leads to reward models that marginalize already underrepresented communities (Ryan et al. 2024). Apart from leading to secondary fairness issues RLHF tried to address in the first place, findings like

---

this point toward an insufficient level of participation and democratic decision-making resulting in reward models that fail to capture the diversity of preferences in heterogeneous populations. LLMs will not become pluralistic and safe in the social sense (Sorensen et al. 2024) unless it is clear what happens if more than one ranking of alternative responses to a prompt exists. In other words, what is done to preserve diversity of preferences with respect to all stakeholders and what type of diversity preserving preference aggregation is used? We do not have satisfactory answers to these questions for commercial models apart from reports on experiments with democratic/collective alignment (e.g., Anthropic 2023c and Eloundou and Lee 2024). Obtaining preference profiles in an accountable, transparent, and responsible manner is difficult and costly, and this is why we see interfaces of commercial chatbots designed in a way that allows real-time collection of customer preferences. Without guarantees on participation and diversity, security practices enacted by RLHF contribute to product development and support of speech acts rather than to closing the gap between the capabilities of the security tool and the desired level of safety guarantees.

Apart from the preference-related defining features, RLHF has also potentially problematic design traits. Lambert et al. (2023) highlighted that since the reward model is based on a pre-trained LLM, issues of the base model could negatively impact the reward model trained on top of it (similar issues impact generative reward modelling where an LLM is used as a judge to determine the ranking of alternatives, Lambert et al. 2024a, p. 17). Krendl Gilbert et al. (2023) suggested that it would be beneficial to maintain reward reports continually updated with information on the model's implementation, its social interfaces, reinforcement learning objectives and the auditing and verification results to simplify keeping track of what we consider design traits and defining features of deployed systems.

Overall, security practices enacted with RLHF show a significant room for improvement mostly with respect to participation and diversity, without which it is difficult to speak about safety guarantees derived from AI alignment.

As the second characteristic of security tools Balzacq et al. (2010) identified actions which are configured by the tools and consist of requirements, procedures and delivery mechanisms. In Sect. 2, it was explained that gaps in alignment left behind by security tools such as RLHF are expected to be found by diverse empirical testing procedures called red teaming. Red teaming practices correspond to actions whose content is 'configured' against the defining features of a particular RLHF run to test the robustness of alignment of the resulting LLM with human preferences.

As shown by Feffer et al. (2024), the adjustability of red teaming practices, which starts with diverse threat models derived from different safety preferences, can render red teaming a mere security theater that is invoked whenever there are concerns over the risks of generative AI on society. The security theater interpretation of red teaming fits well the logic of AI safety securitization, which consists of acknowledging concerns and making speech acts showing that there are actions already underway capable of addressing these concerns (Anthropic 2023d; OpenAI 2023c).

Feffer et al.'s (2024) findings synthesized from existing research on red teaming can be used to critically assess it in terms of requirements, procedures and delivery mechanisms, which constitute actions as the second characteristic of security tools. Since red teaming is a set of evaluation procedures, its results heavily depend on people performing them. In an ideal world, it would be required that selected experts work together with stakeholders to maintain diversity of evaluation by wide participation. In reality, experts are often accompanied by testers from crowd-working platforms, and for them as well as for the experts the question of fair representation and participation remains open (ibid.). Inie et al. (2023) mapped a complex landscape of motivations and goals adopted by people participating in red teaming of LLMs. Diversity of evaluation teams cannot be understated because the ability to find vulnerabilities of the model that unevenly impact different social groups depends on participation. LLM-based red teaming is limited by the number of generated adversarial prompts and the computational budget for the subsequent adversarial training on red teaming data aiming to patch the gaps in the model's alignment (ibid.).

When it comes to testing LLMs on harmful prompts, that is, performing the core of red teaming, Feffer et al. (2024) found evidence of procedures which are neither well-scoped nor -structured. The major contributing factor is the difficulty of meeting participation requirements where instead of diverse stakeholders, crowd-workers test the 'lowest hanging' vulnerabilities and experts represent mostly academe and AI companies (ibid.). It is difficult to comprehensively test versatile LLMs, for which the risk surface can be large and complex with red teams that are biased to focus only on some of its areas. This means that even after using one of the alignment techniques such as RLHF to obtain a supposedly safe LLM, testing procedures are not guaranteed to find all gaps that can be misused. Repeating the process does not necessarily increase the test coverage if the approach to assembling the red team does not change.

Finally, delivery mechanisms, as the last stage of actions configured by security tools, are not without problems either. Despite the existence of open red teaming experiments (Ganguli et al. 2022), detailed reports on the identified vulnerabilities and their mitigation are not always available and there is a lack of standards on the reports' structure, types of disclosed information, and evaluation of the mitigation results (Feffer et al. 2024)[5]. The lack of information complicates the role of stakeholders who cannot submit requests for participation on the basis of ill-conceived or missing tests designed by unrepresentative red teams.

Overall, red teaming procedures following RLHF as actions configured by security tools cannot identify every safety issue left behind by the alignment procedure. The reason for this lies in the fact that to obtain a full test converge of versatile LLMs is close to impossible and also in the fact that the 'configuration' of the action and of its individual stages is often flawed in the same way as the tool itself by lacking in diversity and participation. In this sense, proposals for safe harbors allowing unimpeded safety research of commercial LLMs (Longpre et al. 2024) could become an important tool that not only protects researchers but also increases participation and diversity in AI safety research and contributes to better understanding of the security tools. Generally, a lack of information on the audited AI system always leads to an incomplete picture of the system's risk surface (Casper et al. 2024).

We join the discussion on two remaining Balzacq et al.'s (2010) characteristics of security tools together because public actions and images invoked by security practices enacted with particular tools are closely related. Public actions and images are the end point of (in)securitization processes. In AI safety, security tools such as RLHF, their design traits and defining features, and actions such as red teaming 'configured' by them aim to inspire AI governance (public action) that supports the image of safe and socially responsible innovation in sync with public policies (Anthropic 2023e; OpenAI 2023d).

It is important to note that this process runs in parallel with regulatory efforts that try to ensure society will not be harmed by the gap between the capabilities of AI security tools and the desired level of safety guarantees. Since (in)securitization processes aim to shape public actions and images, it is reasonable to ask whether AI regulation is impacted by security practices of AI companies that develop commercial, versatile LLMs. Red teaming was, for instance, featured in Biden's U.S. Presidential Executive Order on the Safe, Secure, and Trustworthy Development and Use of Artificial Intelligence[6] (rescinded on January 20, 2025) as a vehicle for identifying harmful capabilities of foundation models such as versatile LLMs that need to be mitigated, the results of testing should be then shared with the government. Prior to the Executive Order, there was a public–private partnership between major U.S. companies involved in AI development and the Biden-Harris administration[7]. When looking at the companies' interpretation of the partnership (e.g., OpenAI 2023d), red teaming was the top priority and it turned out to be an important part of the Executive Order as well.

We might ask whether the red teaming's role results from a successful public–private partnership on AI safety or from a securitization process that managed to produce a high-level speech act supported by the security practice. According to Balzacq et al. (2010), tools used to enact security practices reconfigure public action with respect to the image of the threat. We have shown that alignment methods such as RLHF are useful for product development and their use for AI safety does not bring safety guarantees even if accompanied by extensive testing. From a critical perspective, this means that the practice combining alignment procedures with red teaming reconfigured the image of AI safety and paved the way for the public action, e.g., Biden's Executive Order on AI, that accepted LLMs' alignment and red teaming to be sufficient for AI safety.

## 4 From critique to more open, participatory, and sustainable practices in LLM development

In the situation where we lack alternatives capable of delivering genuinely dependable safety guarantees on AI safety, it is perhaps understandable that the securitization process concluded by 'certifying' the product development as a safe practice. Amidst concerns for AI risks of capable systems (Bengio et al. 2023), until (if ever) we have tools capable of deriving dependable AI safety guarantees, more emphasis should be placed on participation and diversity in AI security practices to make sure that they are at least equitable.

One way of achieving this is to consider a different 'baseline'. Alongside large, general-purpose, closed or publicly available, alternatively open source (e.g., Groeneveld et al. 2024), LLMs, multiple small, domain-specific and specialized LLMs could be trained, constrained to perform only a range of domain-specific tasks. There are several benefits to this alternative approach. First, the alignment and red teaming procedures would become more manageable due

---

to a clear and narrow range of tasks that need to be aligned and tested. Second, making the security practice more manageable would also allow developers to engage key stakeholders more effectively, hopefully increasing participation and diversity in the security practices and leading to more accountable sociotechnical AI systems that many call for (e.g., Fazelpour and De-Arteaga 2022). Third, regulatory oversight over specialized, domain-specific LLMs could be performed by government departments or agencies tasked with regulating the given area. This would avoid building parallel (and redundant) capabilities at an AI governance institution specializing in oversight of versatile LLMs.

Concerning participation and diversity in AI security practices, Rauh et al. (2024) identified several problems, among which the model-centered evaluation in combination with automated benchmarks has a potential to obscure social harms that happen outside this narrow context. Relatedly, these evaluations should never be considered value-neutral (ibid.). There are sociotechnical methodologies, such as STAR (Weidinger et al. 2024), aiming to assess impacts on broader social contexts. These contexts are engaged by red teaming instructions that, if the model provided completions, would result in significant safety and social justice violations (ibid.). If methodologies like STAR were added to the LLM development toolbox, increased participation and diversity would become a necessary requirement for red teaming efforts. Without participation and diversity, it would not be possible to determine how to test the model for harmful effects in social contexts that are impossible to understand assuming only the model-centered and supposedly 'value-neutral' perspectives. We will now provide three examples of small, domain-specific models and compare them to large, versatile alternatives.

Currently, it seems reasonable to consider models up to 8 billion parameters (8B) small because they can be used for inference on affordable hardware. For example, Llama-3.1-8B-Instruct, an open-weight (publicly available), instruction-following model from Meta (Grattafiori et al. 2024), can run with decent context lengths even on consumer accelerators (Llama-3.1-8B-Instruct with weights in the half precision floating point format, FP16, and a keys and values cache for the context length of 16 thousand tokens will fit the memory of a high-end consumer GPU, e.g., Nvidia GeForce RTX 4090 with 24 GB of memory, see, e.g., Schmid et al. 2024), though at reduced token throughputs compared to enterprise accelerators. There are many models in this category, some notable examples of open-weight models include: Meta's Llama-3.2-1B & 3B and their instruction-tuned variants developed for edge and mobile devices[8], Microsoft's Phi family of models[9] (although some models from the family have more than 8B parameters, Phi is considered to be a family of small language models), the Gemma family of small models from Google[10], small models from Alibaba's Qwen family[11] (0.5B, 1.5B, 3B, 7B), OpenBMB's MiniCPM family[12] or fully open-source 7B variants of OLMo 2 from Ai2[13].

Llama and Phi models were used to develop small, domain-specific models that were fine-tuned for specific task sets and have, arguably, a more manageable safety profile compared to large models that can be used to perform the tasks as well thanks to their more versatile in-context learning capabilities. LinkedIn developed the EON-8B model based on Llama-3.1-8B-Instruct to suggest matches between candidates and jobs and to generate explanations of the matchings (Bodigutla et al. 2024). Bodigutla et al. (2024) report better or comparable performance to large general models such as GPT-4o or Llama-3-70B and better cost-effectiveness. Alignment and red teaming of an 8B domain-adapted model is still challenging but more manageable compared to large, versatile models. This could be a crucial difference especially in sensitive areas like hiring. Li et al. (2024) domain-adapted several small language models, including Phi-3.5-mini 3.8B (Abdin et al. 2024), for translating natural language tasks to domain-specific code allowing interaction with the hardware fulfilment process that supplies Azure, Microsoft's cloud computing platform. They found that after fine-tuning Phi-3.5-mini ranked best among small models and outperformed large, versatile models like GPT-4-turbo prompted with samples used for adapting the small models (ibid.). Similar to the previous case, the small, domain-adapted Phi-3.5-mini was more cost-effective compared to large versatile models (Li et al. 2024, Table 2) and its alignment arguably more manageable than alignment of a large, general model accessed via in-context learning. Finally, Liu et al. (2023) domain-adapted two versions of Meta's second-generation Llama, 7B and 13B, for chip design (engineering assistant chatbot, electronic design automation script generator, and bug analysis tool). While they found the 13B version best performing, which is slightly above our threshold for small models, the 7B version performed well (ibid.) At individual tasks, the best domain-adapted model was better or matched the performance of a general-purpose, larger model (Llama-2-70B-chat), was more cost-effective (ibid.) and, we might add, its alignment was more manageable. In some cases, as noted by Liu et al. (2023), models' domain adaptation can require investments that are later compensated by lowering operating costs, which are higher for large, general models. Substantive up-front investments are, however, not the rule. Li et al. (2024) estimated that for their use case the cost of adapting the model to the domain was 'negligible'.

Further research is needed to assess the volume of negative externalities, not only in terms of safety and equity but also in terms of environmental impacts (Strubell et al. 2019; Spelda and Stritecky 2020), posed by these two approaches—centralized development of versatile models vs decentralized development of domain-specific, specialized models. There is evidence that during deployment general/multi-purpose models consume more energy (and depending on the power source also possibly produce more carbon emissions) than task-specific (single task) models (Luccioni et al. 2024). We currently lack general evidence on how this relation would look like for versatile vs specialized models, but since the latter will be smaller in the number of parameters, training and runtime inferences could be computationally cheaper and produce fewer negative externalities, all other things being equal. Two of our examples of small, domain-specific models, EON-8B and Phi-3-mini for Azure hardware provisioning, could be used to provide partial evidence. For EON-8B, Bodigutla et al. (2024) reported a lower number of GPUs necessary to support their use case compared to a solution based on a large, general model (see their Fig. 4). This translates into lower computational requirements, energy consumption, and fewer carbon emissions, considering comparable energy sources. Li et al. (2024) compared their domain-adapted Phi-3-mini with 3.8B parameters to the then state-of-the-art large, general model and reported lower costs and higher accuracy, clearly favoring the small model (see their Table 2). In situations like this, running a well-utilized, small, domain-adapted model is bound to be less wasteful than performing domain tasks via in-context learning on large, general models. Costs of adaptation need to be considered as well, but in this case, Li et al. (2024) reported them to be mild. Small and capable base language models could provide a viable, eco-friendly and cost-effective alternative to large, general models in certain domains. Environmental costs of developing small base models are favorable compared to resources involved in the development of large, general models, holding the energy source comparable.

Finally, as explained in Sect. 2, instruction following and generally specialized LLMs are produced using base pre-trained models. It is important to discuss provenance of these base models and practices used to develop them. Fully open models developed with responsible practices (e.g., Groeneveld et al. 2024) should be preferred over commercial or merely publicly available base models (only weights and inference code are available) if full auditability (Casper et al. 2024) and good AI security practices are the top priority.

Genuinely open-source models differ from open-weight models by open access to pretraining data mixtures and training recipes used to develop base models, including data and code required for instruction tuning that produces instruction-following models from base models during post-training. Open-weight models, on the other hand, are available only as weights of the neural network and code that enables using it for completing user-defined tasks. For instance, Ai2's OLMo family of open-source, computationally efficient models offers highly competitive alternatives to open-weight models (see OLMo 2 report, Walsh et al. 2025), for which full data mixtures and detailed training recipes are not publicly available. Open-source artifacts that comprise the OLMo ecosystem, ranging from datasets to models at various sizes, help the scientific understanding of language models development. Specific efforts, such as fully documented and reproducible post-training recipes including the necessary data, e.g., Lambert et al. (2024b), open the stage of LLM development that is crucial for stabilizing, improving and aligning final capabilities of LLMs. Transparency of open-source models is helpful during safety evaluation because open access to pre- and post-training data mixtures and recipes enables identifying unseen tasks that were not part of the model development and, thus, offer a less biased perspective on its safety and alignment. For open-weight and closed models, for which information on pre- and post-training data is not fully disclosed, the degree to which evaluation tasks are unseen (held-out) is not independently verifiable. This can bias the perspective on the models' safety and alignment with respect to user requests.

Open-source and open-weight foundation models come with their own risks caused by the impossibility to control their downstream use. There is an ongoing debate on whether open LLMs pose a greater risk than closed ones accessible only indirectly via APIs, and we briefly touched on this topic when explaining the reversibility of alignment of open LLMs. Qi et al. (2024b) evaluated the robustness of two prominent defenses against fine-tuning/tampering attacks against open models (representation noising, Rosati et al. 2024, and tamper attack resistance, Tamirisa et al. 2024, both methods attempt to unlearn unsafe information from the model so that it is hard to bring it back by reversing alignment of the model) and concluded that they are not robust to what should be inconsequential details of fine-tuning and evaluation pipelines. This means that downstream misuse of open-source and open-weight models remains hard to prevent. Kapoor et al. (2024) constructed a framework for analyzing the marginal risk of open models, i.e., the degree by which a misused open-source or open-weight model increases the social risks over closed models or legacy technologies allowing attackers to perform the task, which could be used to navigate the safety tradeoffs.

## 5 Conclusion

We showed that a securitization process underpinned by a combination of alignment and red teaming procedures reconfigured the public image of AI safety and produced a set of practices that are considered beneficial for safety. This combination of alignment and red teaming is also behind development of LLMs that AI companies use to build products such as chatbots. This situation resulted in regulatory approaches that took closed and commercial, general-purpose LLMs as their baseline, while leaving the status of publicly available or open-source LLMs less clear. Different baselines are possible, such as small, domain-specific LLMs allowing users to perform specialized tasks, on which we demonstrated that AI security practices could be different, more welcoming to participation and diversity—values with which the current AI security practices struggle.

**Acknowledgements** We would like to thank the reviewers for helpful feedback on the manuscript.

**Author contributions** Petr Spelda conceptualized the problem and performed its analysis. Petr Spelda and Vit Stritecky wrote and reviewed the paper.

**Funding** This work was supported by the European Regional Development Fund project "Beyond Security: Role of Conflict in Resilience-Building" (reg. no.: CZ.02.01.01/00/22_008/0004595).

**Data availability** No new data were generated.

## Declarations

**Conflict of interest** On behalf of all authors, the corresponding author states that there is no conflict of interest.